\documentclass[prb,
superscriptaddress,showpacs,amsmath,amssymb]{revtex4}

\usepackage{color}

\begin{document}

\title{Massless graviton in de Sitter as second sound in two-fluid hydrodynamics}

\author{G.E.~Volovik}
\affiliation{Landau Institute for Theoretical Physics, acad. Semyonov av., 1a, 142432,
Chernogolovka, Russia}

\date{\today}

\begin{abstract}
The concept of gravitons and their masses, clear in the case of Minkowski spacetime, remains ambiguous for de Sitter spacetime. Here, we used a two-fluid approach to de Sitter thermodynamics and found a collective mode that is analogous to second sound in the two-fluid dynamics of the de Sitter state. This mode is massless and propagates at the speed of light. This suggests that this second-sound analog is a massless graviton propagating in de Sitter spacetime. The type of graviton this mode represents requires further consideration.
\,
  \end{abstract}
\pacs{
}

\maketitle

 \tableofcontents

\section{Introduction}

The de Sitter spacetime is the unique maximally symmetric solution to the Einstein equation with a positive cosmological constant.\cite{Gazeau2023}  The de Sitter state has also the unique thermodynamics properties.\cite{Volovik2025,Volovik2024} 
The de Sitter spacetime has the cosmological horizon, which is similar to the black hole horizon, and thus it can be described by thermodynamics related to horizon. 
However, as distinct from the black hole, which is the compact object, the de Sitter state is homogeneous: all the points of de Sitter state (inside and outside horizon) are equivalent.
The observer at any point of de Sitter space (inside and outside horizon) feels the same local temperature $T=H/\pi$, where $H$ is the Hubble parameter (see also \cite{Maxfield2022}). This suggests that the de Sitter state can be also described by the local thermodynamics, which is characterized by the entropy density (see also \cite{Diakonov2025}). This is supported by the holographic connection between the entropy of the Hubble volume and the horizon entropy: the entropy density integrated over the Hubble volume gives exactly the Gibbons-Hawking entropy of the cosmological horizon. 

Being the homogeneous state, the de Sitter Universe has the signatures of the  so-called Kronecker anomaly. \cite{PolyakovPopov2022} The consequence of the Kronecker anomaly for thermodynamics is that thermodynamics may contain the terms that are absent in dynamic equations. In the de Sitter state,  the first law of de Sitter thermodynamics contains the extra term $K{\cal R}$, where $K=1/16\pi G$ is the gravitational coupling and ${\cal R}$ is the scalar curvature. Due to Kronecker anomaly, the homogeneous curvature ${\cal R}$ in the de Sitter thermodynamics behaves as matter. This makes it possible the heat exchange without ordinary matter. If the ordinary matter is introduced into the de Sitter environment, the heat exchange between the Sitter heat bath and the ordinary matter leads to the decay of the de Sitter state towards the Minkowski vacuum.

Here we consider another consequence of the unique properties of the de Sitter state, which
resembles the collective behavior of multiparticle systems described by thermodynamics and hydrodynamics. Thermodynamics and hydrodynamics are two cornerstones that also apply to relativistic systems and their quantum vacua. The two-fluid hydrodynamics introduced by Landau for superfluid $^4$He is an example of the collective behaviour, which is also applicable to the relativistic systems. It describes the thermodynamics and dynamics of the de Sitter state.\cite{Volovik2025} Within the two-fluid hydrodynamics there is the so-called second sound, which is the propagating mode of the entropy density. In the two-fluid cosmological hydrodynamics of the de Sitter state, this corresponds to the massless graviton propagating with the speed of light.

In Section \ref{TwoComponents} the main elements of the two-fluid hydrodynamics of the de Sitter state are introduced. The analog of the coherent (superfluid) component of the cosmological liquid is considered  in Sec. \ref{VacComponent}. It is represented by the $\Lambda$-term -- the vacuum energy (dark energy). The analogue of the normal component is considered in Sec. \ref{GravComponent}. The behaviour of this thermal component follows from the observation that matter immersed in the de Sitter vacuum feels this vacuum as a heat bath with temperature $T=H/\pi$, where $H$ is the Hubble parameter, see also Ref.\cite{Maxfield2022}. This heat bath, which represents the thermal component of the cosmological fluid, comes from the gravitational degrees of freedom (the scalar curvature). The thermal component behaves as the Zel’dovich stiff matter, and the speed of sound propagating in this component equals the speed of light, $s^2=dP/d\rho=c^2$.

In Sec. \ref{EquilSec} the equilibrium state of the de Sitter state described by the two-fluid hydrodynamics is considered with the thermodynamic relations between the normal and superfluid components. 
In Section \ref{SecondSoundSec}, it is shown that this Zel'dovich type sound mode actually represents the second sound in the Landau two-fluid hydrodynamics. Its velocity $s_2$ is obtained from the Landau equation for the spectrum of the second sound applied to the de Sitter cosmological fluid. This equation demonstrates that second sound is the gapless mode propagating with the speed of light, $s_2=c=s$.

In the \ref{SecDiscussion} section, we attempt to identify this collective mode as a type of graviton propagating in the de Sitter background. A possible hypothesis is a massless graviton with zero helicity.

\section{Two components of de Sitter hydrodynamics}
\label{TwoComponents}

\subsection{Vacuum component}
\label{VacComponent}

Let us recall the basic elements of the two-fluid hydrodynamics in superfluids and in the de Sitter state of quantum vacuum.\cite{Volovik2025} 
We start from the quantum vacuum state -- Minkowski vacuum. The condensed matter analog of the this state is the ground state of the superfluid liquid without excited quasiparticles representing matter. Since the temperature of such quantum vacuum is zero, $T=0$,  there is the following thermodynamic Gibbs-Duhem relation in superfluid liquid such as bosonic superfluid $^4$He:
 \begin{equation}
 \epsilon(n) - \mu n=-P\,.
\label{GS4He}
\end{equation}
Here $n$ is the density of the $^4$He atoms; $ \epsilon(n)$ is the energy density; $\mu=d\epsilon/dn$ is the chemical potential; and $P$ is pressure. 

We do not know what are the "atoms of the vacuum", but we know that the laws of the macroscopic thermodynamics are general. They do not depend on the micro-structure and on the types of atoms, and thus are equally applied to the relativistic and non-relativistic systems. For example,  to describe the vacuum energy, one may use the 4-form field $q$ introduced by Hawking\cite{Hawking1984,Duff+Wu}
 as the analog of the variable $n$.  In this the so-called $q$-theory,\cite{KlinkhamerVolovik2008}  the Gibbs-Duhem relation has the form similar to that in Eq.(\ref{GS4He}) with $\mu=d\epsilon/dq$:
 \begin{equation}
 \epsilon(q) - \mu q=-P\,.
\label{GSq}
\end{equation}
It is important that in the corresponding Einstein equations, the cosmological constant $\Lambda$ (or dark energy) enters as 
 \begin{equation}
\Lambda\equiv \epsilon_{\rm vac}= \epsilon(q) - \mu q=-P\equiv -P_{\rm vac}\,.
\label{Lambda}
\end{equation}
That is why one obtains the equation of state $w=-1$ for the dark energy:
 \begin{equation}
 \epsilon_{\rm vac}= - P_{\rm vac}\,.
\label{EqStates}
\end{equation}
The same equation of state takes place for the ground state of superfluid $^4$He. The reason for that is that the thermodynamic energy of this quantum liquid is $E=<{\hat H}>-\mu <{\hat N}>$, where ${\hat N}$ is the the particle number operator.

Another important property is the nullification of the vacuum energy in the absence of the external environment and thus in the absence of the external pressure, $\epsilon_{\rm vac}= - P_{\rm vac}=-P =0$. This is well known for the self-sustained superfluid $^4$He, which remains stable without the external pressure. That is why if the Minkowski vacuum is also self sustained, its vacuum energy is automatically zero, irrespective of its microscopic structure. 

This also demonstrates that the representation of the vacuum energy in terms of the zero-point energies of the collective modes is misleading. The collective modes are the low energy excitations of the quantum vacuum and they cannot be responsible for the micro-structure of the deep vacuum. Thus, the cosmological constant problem is resolved by the thermodynamic law in Eqs. (\ref{GS4He}) and (\ref{GSq}). In the full equilibrium at $T=0$ and in the absence of external pressure, the large contributions $\epsilon(n)$ and $\epsilon(q)$ are fully compensated by the contributions of the corresponding chemical potentials $\mu$. Both $\epsilon(n)$ and $\epsilon(q)$ describe the micro-physics (physics of the $^4$He atoms and the ultraviolet physics of the "atoms of the vacuum"), which cannot be obtained by the quantization of the low-energy (infrared) collective modes (phonons and photons).

\subsection{Gravitational component}
\label{GravComponent}

The non-zero value of the cosmological constant is obtained if the equilibrium state is disturbed either by perturbations of the vacuum state or at $T\neq 0$. The perturbations and thermal excitations play the role of the normal component of the system. Here we consider the analog of  normal component in the de Sitter state,  which is played by the gravitational degrees of freedom. The thermal properties of the de Sitter state follow from the observation that any object (or observer) immersed in the de Sitter environment feels this environment as a heat bath with temperature $T=H/\pi$, where $H$ is the Hubble parameter. 
Then, from the free energy density of the gravitational component\cite{Volovik2025d}  one obtains the entropy density of the de Sitter state, $S= - dF/dT$,
see Eq.(\ref{LocalEnergyDensity}). This gives
 \begin{equation}
 S=\frac{3\pi T}{4G}\,.
\label{EntropyDensity}
\end{equation}

Entropy density $S$ and temperature $T$ are the local thermodynamics variables of the de Sitter state, but they are formed due to existence of the cosmological horizon and they are also responsible for the Gibbons-Hawking entropy $S_{\rm GH}$ related to cosmological horizon. The latter is obtained as the entropy of the Hubble volume $V_H$ (the volume inside the cosmological horizon): \begin{equation}
S_{\rm GH}= SV_H=\frac{A}{4G}\,,
\label{GibbonsHawking}
\end{equation}
where $A$ is the horizon area. This is valid for space dimension $d=3$, while it is shown\cite{Volovik2025b} that for the other space dimensions, the horizon entropy deviates from its Gibbons-Hawking value
\begin{equation}
SV_H=\frac{(d-1)A}{8G}=\frac{d-1}{2}S_{\rm GH}\,.
\label{GibbonsHawking2}
\end{equation}
The local thermodynamics of de Sitter states has been also considered by Diakonov in Ref. \cite{Diakonov2025}. However, with his approach, the local temperature is equal to the Gibbons-Hawking temperature $T_{\rm GH}=H/2\pi$.

The important property of the local entropy is that one has the following relation between this entropy and the thermodynamic quantity related to the gravitational degrees of freedom,  which is obtained from the Einstein equations:
\begin{equation}
ST= K{\cal R}\,.
\label{EntropyCurvature}
\end{equation}
Here ${\cal R}=12H^2$ is the scalar curvature, and $K=df/d{\cal R}$ is its thermodynamically conjugate quantity in the $f({\cal R})$ theory. In the conventional Einstein action it is the gravitational coupling,
 $K=\frac{1}{16\pi G}$. This connection demonstrates that the role of the thermal normal component of the cosmological fluid is played by the gravitational degrees of freedom of the Sitter state, which are represented by the scalar curvature ${\cal R}$. 
 
 The emergence of the thermodynamic variable ${\cal R}$, which describes the gravitational degrees of freedom and forms the normal component of the cosmological two-fluid hydrodynamics, is the result  of the so-called Kronecker anomaly \cite{PolyakovPopov2022}, or of the Larkin-Pikin effect \cite{LarkinPikin1969}. In this anomaly, the extra degrees of freedom emerge in the fully homogeneous state, i.e. at ${\bf k}=0$. These extra parameters are space-independent but participate in thermodynamics, as it happens with the curvature  ${\cal R}$ in de Sitter. 
 
 One can write the conventional Gibbs-Duhem thermodynamic relation for the gravitational normal component with its energy density $\epsilon_n$ and partial pressure $P_n$:
 \begin{equation}
\epsilon_n - TS= - P_n\,.
\label{GDgrav}
\end{equation}
Then using Eq.(\ref{EntropyCurvature}) and the condition for the absence of external pressure, $P_n + P_{\rm vac}=0$ (see Section \ref{EquilSec}) one obtains $\epsilon_n=\epsilon_{\rm vac}=-P_{\rm vac}=P_n$.
So, the cosmological normal component behaves similar to that of the Zel'dovich stiff matter \cite{Zeldovich1962} with equation of state $w_n=1$, while for superfluid (dark energy) component the equation of state is $w_s=-1$.

This thermal gravitational component is responsible for the local de Sitter entropy. From the local energy density of the gravitational component
 \begin{equation}
\epsilon_n = \epsilon_{\rm vac}=\frac{3}{8\pi} \frac{H^2}{G}=\frac{3\pi}{8} \frac{T^2}{G}\,,
\label{LocalEnergyDensity}
\end{equation}
one obtains the entropy density of the de Sitter state in Eq.(\ref{EntropyDensity}), $S=-dF/dT=d\epsilon_n/dT= 3\pi T/4G$.

Since the thermal gravitational component behaves as the Zel'dovich stiff matter, one obtains that there are the sound waves propagating with the speed of light $s^2=dP_n/d\epsilon_n= c^2$.
In Section \ref{SecondSoundSec} we will demonstrate that these gravitational waves correspond to the second sound in the two-fluid de Sitter thermodynamics.

\section{Equilibrium de Sitter state and its second sound}
\label{EquilSec}

\subsection{Equilibrium state}
\label{EquilSec}

To consider the equilibrium state of the two-fluid de Sitter hydrodynamics, let us  discuss the pressure arising from gravitational degrees of freedom.
The standard Hilbert stress-energy tensor, defined as the functional derivative of matter part of the Einstein-Hilbert action, does not contain this pressure. This is an example  when the conventional approach to the stress-energy does not work. Another example is the introduction of the Landau-Lifshitz pseudotensor.\cite{LandauLifshitz} In our case the reason is the Kronecker anomaly, \cite{PolyakovPopov2022} which takes place for some homogeneous thermodynamic states.  The de Sitter state belongs to this class, because it is invariant under the special type of translations: in the Painleve-Gullstrand coordinates with the shift vector ${\bf v}({\bf r})=H{\bf r}$ the metric is invariant under the combined translations  ${\bf r} \rightarrow {\bf r} + {\bf r}_0 e^{Ht}$, see Eq.(\ref{PG}). Here, the position ${\bf r}_0$ is arbitrary, it can be inside and outside the horizon. This transformation demonstrates that all the points of the de Sitter vacuum, both inside and outside the horizon, are equivalent and have the same scalar curvature. The homogeneous curvature provides the contribution to the Einstein action which, being a derivative, does not affect Einstein's equations. But in the de Sitter state this term contributes to the thermodynamics energy, giving rise to $\epsilon_n$.

If there is no external pressure, the partial pressures of the two components (negative vacuum pressure and positive pressure due to curvature) cancel each other, $P_n + P_{\rm vac}=0$, and as a result one obtains $\epsilon_n = \epsilon_{\rm vac}$. The zero  external pressure condition is imposed well outside the horizon. For example, we can consider the de Sitter ball with radius much larger than the horizon radius, $R_{\rm ball} \gg 1/H$. This radius should be large enough to avoid the pressure caused by the surface tension of the ball. Then one has $P_n + P_{\rm vac}=0$.

So, the thermodynamic quantities characterizing the equilibrium state of de Sitter Universe in the two-fluid cosmology (these are: densities of the "superfluid" and "normal" components; their energy densities and partial pressures; the entropy density and temperature of the "normal" component) obey the following relations:
 \begin{equation}
 \epsilon_s \equiv \rho_s c^2= -P_s=\Lambda \,\,,\,\, w_s=-1 \,,
\label{EnergyDensityS}
\end{equation}
 \begin{equation}
 \epsilon_n \equiv \rho_n c^2=TS - P_n=K{\cal R} - P_n=P_n= \frac{1}{2} ST \,\,,\,\, w_n=1\,.
\label{EnergyDensityN}
\end{equation}
Here $\epsilon_s$ and $\epsilon_n$ are the energy densities of superfluid and normal components; 
$\rho_s$ and $\rho_n$ are the corresponding superfluid and normal densities; $P_s$ and $P_n$ are their partial pressures with $P_s+P_n=0$ in the absence of external environment; and $c$ is speed of light.
 All this demonstrates that in this de Sitter hydrodynamics/thermodynamics, the density of the superfluid component  (dark energy) and the density of the normal component (gravitational analog of stiff matter) have equal values, $\rho_s= \rho_n$. 
 
  Eqs.(\ref{EnergyDensityS}) and (\ref{EnergyDensityN}) remain valid in the $f({\cal R})$ gravity,
\cite{Volovik2025c} where $K=df/d{\cal R}$. This is also true for any other type of modified gravity, as long as it admits the de Sitter solution. The reason is that in the de Sitter state all curvature tensors are expressed in terms of the scalar curvature ${\cal R}$, which is the main element of the two-fluid hydrodynamics of the de Sitter vacuum.

It should be noted that here we do not take into account the instability and decay of the de Sitter state. A pure de Sitter state does not decay. The de Sitter state becomes unstable upon the addition of ordinary matter. The latter breaks de Sitter symmetry, resulting in the energy exchange between the gravitational normal component and the conventional matter component, leading to the decay of the de Sitter state  toward the Minkowski vacuum with $\rho_s= \rho_n=0$.\cite{Volovik2025} This suggests a natural solution to the cosmological constant problem.

\subsection{Second sound in de Sitter hydrodynamics}
\label{SecondSoundSec}

Let us consider a particular consequence of the connection between cosmological two-fluid hydrodynamics and ordinary two-fluid hydrodynamics, represented, for example, by superfluid $^4$He. We are interested in the analog of the second sound -- the propagating waves of the entropy density $S$, which is concentrated in the thermal normal components of the conventional and cosmological fluids. 

The velocity $s_2$ of the second sound in the two-fluid hydrodynamics has the following general form, see e.g. \cite{Schmitt2014}:
\begin{equation}
s_2^2=\frac{TS^2}{C_V\rho} \,\frac{\rho_s}{\rho_n} \,,
\label{EntropyRatio}
\end{equation}
where $C_V$ is the specific heat; and  $\rho$ is the liquid density, which in the cosmological case is \begin{equation}
\rho =\rho_n +\rho_s=2\rho_n
 \,,
\label{density}
\end{equation}
according to Eqs. (\ref{EnergyDensityS}) and (\ref{EnergyDensityN}).

The thermodynamics of the cosmological normal component is similar to that of Zel'dovich stiff matter, that is why the entropy density $S$ in Eq.(\ref{EntropyDensity}) is linear in temperature. As a results,  one has $C_V=S$, and the speed of the analogue of the second sound is:
\begin{equation}
s_2^2= \frac{TS}{2\rho_n}\,.
\label{EntropyRatio2}
\end{equation}
Then from Eqs. (\ref{EntropyRatio2}) and (\ref{EnergyDensityN}) one obtains that the velocity of the second sound coincides with the speed of light:
\begin{equation}
 s_2=c \,.
\label{SpeedSecondSound}
\end{equation}

  This is natural, since the second sound in this cosmological two-fluid thermodynamics is the mode propagating in the thermal gravitational component. Since this component behaves as the Zel'dovich stiff matter, one has
\begin{equation}
  s_2^2=s^2=\frac{dP_n}{d\rho_n}=w_nc^2=c^2\,.
\label{SpeedSound}
\end{equation}

At first glance this seems strange: why would Landau's classical two-fluid hydrodynamics equations lead to a mode that propagates at the speed of light. But apparently the two-fluid approach is quite general. It follows from the general laws of thermodynamics, which apply also to the relativistic quantum vacuum. In particular, thermodynamics allows us to solve the cosmological constant problem.\cite{Volovik2025} The mechanism of cancellation of the vacuum energy in the ground state of the system is purely thermodynamic and does not depend on whether the vacuum is relativistic or not. The relativistic behaviour of the cosmological second sound  follows directly from the Landau equation for the second sound, if the parameters of de Sitter two-fluid thermodynamics are used. This supports the applicability of the two-fluid approach to the de Sitter state.

\section{Discussion. What are collective modes of de Sitter state?}
\label{SecDiscussion}

The analogue of the second sound mode comes from the dynamics of the normal component of the cosmological fluid represented by scalar curvature ${\cal R}$. That is why it is a kind of graviton, a perturbation of scalar curvature  ${\cal R}$, which exists in de Sitter in addition to the usual two graviton modes. This mode propagates in a completely homogeneous de Sitter state, in which the local temperature and entropy density are space independent. The equation (\ref{SpeedSound}) for the speed of sound is derived in the limit of the large wave length, which in the de Sitter state  exceeds the radius of the cosmological horizon, i.e. when the second sound is the superhorizon mode.  If this mode is not an artefact of analogy, then what type of gravitational wave does this second sound belong to? 

There are several options: the second sound in de Sitter spacetime may correspond to the spin-0 graviton,\cite{Wang2011} to helicity-0 graviton,\cite{Izumi2009} to the pseudo Nambu-Goldstone bosons emerging in the slow-roll limit of inflation,\cite{Freese1990,Kitazawa2023,Merchand2025}
 or to the soft graviton mode viewed as the Goldstone mode of the spontaneously broken symmetry in de Sitter space.\cite{Sloth2025}  It is possible that the gravitational ${\cal R}$ mode (or the $\zeta$ mode \cite{Maldacena2002,Sloth2025}) represents the second sound in the cosmological two-fluid hydrodynamics.

De Sitter state violates (or breaks) translational symmetry. In the Painleve-Gullstrand form
\begin{equation}
  ds^2=-c^2 dt^2 + (d{\bf r}-H{\bf r}dt)^2\,,
\label{PG}
\end{equation}
 instead of translations in space and time one has symmetry under the space translations combined with time,
 ${\bf r}\rightarrow {\bf r}+ {\bf r}_0e^{Ht}$. That is why one symmetry element is broken, and this should lead to a single Goldstone mode. Since the entropy density is the local thermodynamic variable, its dynamics may correspond to this Goldstone boson. It should also be noted that there is still no consensus regarding the mass of modes in the de Sitter background, see e.g. \cite{Garidi2003,Akhmedov2017,Gazeau2023,Sadekov2024,Hinterbichler2025}. This is why what is massive in Minkowski spacetime may appear massless in de Sitter spacetime.
  
 Note that in this cosmological two-fluid hydrodynamic, the first sound is absent. This is explained by the fact that the energy density of the "superfluid" component $\Lambda$ is considered here as a constant (cosmological constant). In case of dynamical vacuum energy, the corresponding collective mode is the analogue of the Higgs mode, see e.g. the so-called $q$-theory,\cite{KlinkhamerVolovik2008} where the vacuum energy is described in terms of the 4-form field introduced by Hawking.\cite{Hawking1984,Duff+Wu}
 The same oscillatory behaviour takes place in Starobinsky inflation.\cite{Starobinsky1980}
 
Magnetic field also contributes to the Kronecker anomaly.  In cosmology, the homogeneous magnetic field is possible in the presence of the tuned cosmological constant, 
the so-called Bonnor–Melvin-$\Lambda$ Universe, see e.g Refs.\cite{Plebanski1979,Astorino2012,Zofka2019,Ahmed2024,Castro2024}. It would be interesting to study the thermodynamics of such state and the collective modes therein.

\section{Conclusion}

The Landau two-fluid hydrodynamics properly describes the thermodynamics and dynamics of the de Sitter state. 
Two components of the de Sitter cosmological fluid come from the two independent contributions to the Einstein-Hilbert action: the  scalar curvature term and the dark energy term. The latter corresponds to the superfluid component of the cosmological fluid with equation of state $w=-1$, while the former represents the normal component, which is responsible for the thermal properties of the de Sitter state with its local temperature and entropy density. Its equation of state is $w=1$, which is similar to the equation of state of the Zel'dovich stiff matter. The entropy density of this normal component integrated over the Hubble volume reproduces the Gibbons-Hawking entropy associated with the cosmological horizon.

 One of the consequences of the  two-fluid hydrodynamics is the existence of the second sound mode -- the propagating waves of entropy density. In the cosmological two-fluid dynamics, where the normal component is represented by the scalar curvature,  the second sound mode is a type of graviton propagating in the de Sitter state. From the Landau equation for the speed of the second sound it follows that, although the Landau equation is classical, with the parameters of de Sitter it gives the relativistic result. This graviton is massless and propagates  with the speed of light. This demonstrates the universality of the Landau two-fluid model and could provide clues to understanding the spectrum of gravitons in the de Sitter universe, a topic that still has many unexplored subtleties.

\end{document}